\documentstyle[aps,pre,twocolumn,graphics]{revtex}
\begin{document}
\draft

\twocolumn[
\hsize\textwidth\columnwidth\hsize\csname @twocolumnfalse\endcsname

\title{Globally coupled maps with asynchronous updating}
\author{Guillermo Abramson\\
{\em Max-Planck-Institut f\"ur Physik komplexer Systeme\\
N\"othnitzer Strasse 38, D-01187 Dresden, Germany}}
\author{ Dami\'an H. Zanette\\
{\em Consejo Nacional de Investigaciones Cient\'{\i}ficas y
T\'ecnicas \\ Centro At\'omico Bariloche and Instituto Balseiro,
8400 Bariloche, Argentina}}

\maketitle
\widetext
\advance\leftskip by 57pt
\advance\rightskip by 57pt

\begin{abstract}

We  analyze  a   system  of  globally   coupled  logistic  maps   with
asynchronous updating. We show that its dynamics differs  considerably
from that of the synchronous case. For growing values of the  coupling
intensity, an  inverse bifurcation  cascade replaces  the structure  of
clusters  and  ordering  in  the  phase  diagram. We present numerical
simulations  and  an  analytical  description  based  on  an effective
single-element  dynamics  affected  by  internal fluctuations. Both of
them show how global coupling is able to suppress the complexity of the
single-element evolution. We  find that, in  contrast to systems  with
synchronous update,  internal fluctuations  satisfy the  law of  large
numbers.

\end{abstract}

\pacs{PACS: 05.45+b, 05.90+m}

]
\narrowtext

\section{Introduction}

Introduced by  Kaneko in  1984 \cite{kaneko84a},  coupled map lattices
have proved to be a powerful tool in the study of spatiotemporal chaos
and       pattern       formation       in       complex       systems
\cite{kaneko89a,kaneko90a,chate92a,chate92b,chate92c,perez93a}.   They
have found  applications as  models in  different branches  of science
\cite{wang93a,dominguez93a,heagy94a,marcq97a}.  Globally  coupled maps
\cite{kaneko90a}, which bear similarity to the Sherrington-Kirkpatrick
model,  constitute  a  kind  of  mean-field  extension  of coupled map
lattices.  In an ensemble  of globally coupled maps, the  elements are
seen  to  form  clusters  of  synchronized  activity.  As the coupling
increases,  a  variety  of  phases  can be identified---from coherent,
through ordered and partially ordered, to turbulent---depending on the
number of  clusters in  the system.   For sufficiently  large coupling
intensities,  the  generic  behavior  consists of full synchronization
of  the  whole  population,  with  all  the  elements having identical
evolution.

In usual  models of  globally coupled  maps the  elements update their
state at the same time, namely, their evolution is synchronous. For an
ensemble of  $N$ elements  whose individual  dynamics is  given by the
nonlinear  map  $x_i(t+1)=f[x_i(t)]$,  a  typical  scheme  for  global
coupling is given by
\begin{equation}
x_i(t+1)=(1-\epsilon)f[x_i(t)]+\frac{\epsilon}{N}\sum_{j=1}^{N}
f[x_j(t)], \  i=1,\dots ,N,
\label{sync}
\end{equation}
where  $\epsilon\in  [0,1]$  is   the  coupling  intensity.   Equation
(\ref{sync}) is applied synchronously to  all the maps of the  system,
with the values of all the sites at the previous time as inputs.

From a  realistic viewpoint,  however, synchronous  updating does  not
seem   to   be   very   plausible   in   models   of   real    systems
{\cite{gade95a,perez96a,harvey97a,rolf98a}.    As    discussed,    for
instance, for neural networks \cite{hertz91}, an independent choice of
the times at which the elements of a given complex system update their
respective states  should provide  a closer  approximation to reality.
It has however to be  mentioned that, in systems with  local coupling,
asynchronous  update  has  been  shown  to  lead  to  trivial behavior
\cite{chate92c}.  We   present  here   a  first   analysis  of    {\em
asynchronous}  globally  coupled  maps  and  find their behavior to be
completely different from  that of the  usual synchronous models.   We
have analyzed two possible asynchronous schemes:
\begin{description}
\item[(a)] At  each time  step, update  the elements  according to the
sequence given by  a prescribed ``list''  that contains every  element
once. The order  of the elements  in the list  is chosen at  random at
each step.
\item[(b)] Divide each time step  into $N$ substeps. At each  substep,
update  one  element  chosen  at  random.  After  $N$ substeps the map
operates, on average, once over each element in the system.
\end{description}

For the sake  of clarity, let  us formulate mathematically  the second
scheme. Let $\eta(t)$  be a stochastic  process that takes  an integer
value $\eta=1,2, \dots, N$ at each substep, with uniform  probability.
The dynamics proceeds in the following way:
\begin{equation}
x_i(t+\frac{1}{N})=\left\{ \begin{array}{ll}
                   (1-\epsilon)f[x_i(t)]+{\displaystyle \frac{\epsilon}{N}
                   \sum_j} f[x_j(t)] &
                   \mbox{if $i=\eta(t)$,} \\
		   x_i(t) & \mbox{if $i \neq \eta(t)$.}
		   \end{array}
	   \right.
\label{async}
\end{equation}
In the first scheme, $\eta(t)$ takes the values $1,2,...,N$ at random,
but only  once within  each time  step. It  can be  seen that, in this
dynamics, the mean  field  $F(t)=N^{-1} \sum_jf[x_j(t)]$ has different
values when acting over different elements, as time proceeds by  steps
of length $1/N$. This feature and the fact that asynchronous  updating
incorporates a  stochastic component  in the  evolution, contrast with
the deterministic synchronous scheme (\ref{sync}).

Evolution under the two asynchronous schemes introduced above displays
the same global properties. However, in the second scheme one or  more
elements may fail to be updated  in some time step. As a  consequence,
since $N$  updates have  to be  made during  a time  step, some  other
elements  will  be  updated  more  than  once.   This  has  nontrivial
consequences  in  the  dynamics,  to  be discussed below. Besides, the
second scheme  has the  advantage of  being much  faster computationally.

In this  paper, we  analyze an  ensemble of  globally coupled logistic
maps.   We  show  that   the  dynamics  under  asynchronous   updating
substantially differs from that of the synchronous case. The structure
of clustered phases  completely dissapears. In  its place, an  inverse
bifurcation cascade develops as  the coupling intensity is  increased.
In  the  following,   we  analyze  this   phenomenon  from   numerical
simulations and construct  a phase diagram  for the different  regimes
displayed by the dynamics.  Then, we propose an analytical explanation
for the bifurcation cascade,  whose results compare successfully  with
the  numerical  data.   Finally,  we  discuss  how  the   fluctuations
generated  in  the  internal  dynamics  are  reflected  in the average
evolution.

\begin{figure}[t]
\centerline{\resizebox{\columnwidth}{!}{\includegraphics{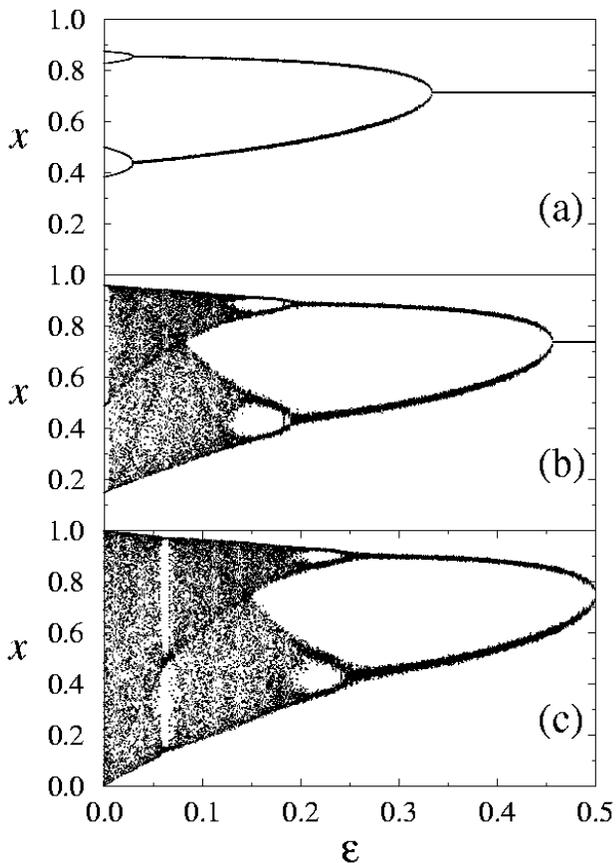}}}
\caption{Bifurcation diagrams of a  system of $N=1000$ logistic  maps,
as a function of the  coupling intensity $\epsilon$. The diagrams  are
constructed  following  the  evolution  of  a  single element taken at
random, with (a) $\lambda=3.5$,   (b) $\lambda =3.84$, (c)  $\lambda =
4$.}
\label{bif}
\end{figure}

\section{Inverse bifurcation cascade and phase diagram}

In  this  section  we  present  extensive numerical simulations of the
evolution given in  Eq. (\ref{async}) for  the standard logistic  map,
$f(x)= \lambda  x(1-x)$.   We concentrate  in the  values of $\lambda$
where  this  map  displays  its  bifurcation  cascade, $3<\lambda <4$,
leading through period doubling from  a stable fixed-point state to  a
completely developed  chaotic regime.   For reasons  that will  become
evident later, we  restrict the coupling  intensity $\epsilon$ to  the
interval $0<\epsilon<0.5$.

In Fig. \ref{bif} we show  the bifurcation diagrams for the  evolution
of a  single element,  chosen at  random from  a population  of $10^3$
maps, as  a function  of $\epsilon$,  for three  values of  $\lambda$.
These  diagrams  are  constructed  as  usual.  For  given  values   of
$\epsilon$ and $\lambda$ the system is let to evolve until transients
have elapsed. Then,  the state of  the chosen element  is recorded and
plotted at some successive time steps.

Figure \ref{bif}(a) shows  the bifurcation diagram  for $\lambda=3.5$,
where the (uncoupled)  logistic map evolves  in a period-4  orbit.  As
$\epsilon$ grows, the four initial branches collapse into two branches
which, in turn,  merge into a  single branch. Increasing  the coupling
intensity leads thus the evolution  of a single element to  display an
inverse bifurcation cascade. The same scenario occurs for other values
of $\lambda$.   For $\lambda=0.38$, where  the logistic map  is within
its largest period-3 stability window, increasing $\epsilon$ leads the
evolution to successively exhibit chaotic regimes of one and two bands
and, eventually, stable branches  that finally collapse into  a single
stable  state  (Fig.  \ref{bif}(b)).  For  $\lambda=4$, in the extreme
chaotic regime of  the logistic map,  the full bifurcation  diagram is
run over backwards (Fig. \ref{bif}(c)).

It is apparent  from Fig.   \ref{bif} that the  evolution of a  single
element  in the ensemble is  subject to  the action of  noise. Indeed,
the   branches   of   nonchaotic    evolution   are   not    perfectly
defined---except in the case of a single stable state---and high-order
bifurcations, close to   the onset of  chaos, are clearly  suppressed.
This noise, which is   produced internally  in   the system  via   the
asynchronous  updating, induces spreading of the states visited during
the evolution when the coupling is different from zero.

The branches of nonchaotic evolution, with their small but  noticeable
dispersion, are  what remains  of the  stable periodic  orbits of  the
deterministic map under  the effect of  the internal noise.   In these
branches, the  evolution driven  by the  two schemes  presented in the
Introduction  differ.  Scheme  {\bf(a)}  defines  true  noisy periodic
orbits,  where  each  element  visits  the  set  of  available  states
following the same periodic sequence as the deterministic map.  On the
contrary, scheme {\bf(b)} allows some elements to skip the update in a
time step.  Consequently, some  other elements  are updated  more than
once.  As  a result, the  observed successive states  do not follow  a
periodic sequence.   To simplify the  discussion, in the  following we
will call these regimes of noisy nonchaotic (periodic or  nonperiodic)
evolution ``period-$n$'' orbits, referring rather to the set of  states
visited by each element.

The chaotic bands, meanwhile, are also blurred by the internal 
noise. We will  thus  refer to a regime where a noise-free map would
exhibit deterministic chaotic evolution as an  ``$n$-band''  chaotic 
region.

To summarize the variety of behavior observed for a single element  as
$\epsilon$ and $\lambda$ vary, we have constructed a phase diagram  in
the plane spanned  by these two  parameters for a population of $10^3$
elements. For each value of $\epsilon$ and $\lambda$ the evolution has
been calculated with scheme {\bf (b)}  during $10^4$ time steps. After
a transient of $5\times 10^3$  steps, a 500-column histogram over  the
states visited  by a  single element  in the  remaining $5\times 10^3$
steps  has   been  produced   to  identify   the  kind   of  evolution
corresponding to those values of the parameters. The phase diagram  is
shown in Fig.  \ref{phases}.  Each region in this diagram  corresponds
to a different kind of attractor.

\begin{figure}[t]
\centerline{
\resizebox{\columnwidth}{!}{\rotatebox{-90}{\includegraphics{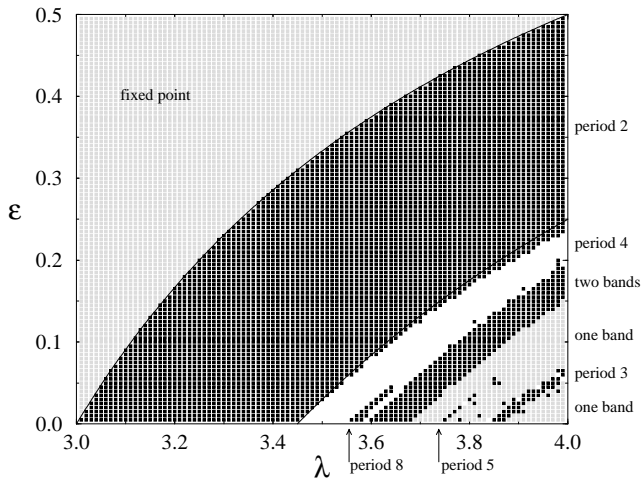}}}}
\caption{Phase diagram of a system of $N=1000$ logistic maps. The  two
lines are  the analytic  curves given  by Eqs. (\protect\ref{ecritic})
and (\protect\ref{ecritic2}).}
\label{phases}
\end{figure}

The large  upper-left region  corresponds to  a fixed  point.  In this
zone  of  the  parameter  space,  from  any initial condition, all the
elements  are  attracted  to  the  same  stable  fixed point.  Without
coupling---i.e.  for  $\epsilon=0$,  on  the  horizontal  axis  of the
plot---such behavior would be observed for $\lambda<3$ only. At larger
values  of  $\lambda$  and,  notably,  even  up to the fully developed
chaotic regime ($\lambda=4$), there always exists a coupling intensity
able to  suppress the  complex behavior  of the  uncoupled system.  As
mentioned above,  in this  fixed-point regime  internal noise  is also
suppressed.

Below  the  fixed-point  region,  there  is  a  series of zones shown,
alternately, with black squares and empty space in Fig.  \ref{phases}.
In this phases the  evolution of single elements  display ``periodic''
and ``chaotic''  noisy orbits,  as indicated  in the  plot. Due to the
noise, the limits  between these regions  are not sharply  defined and
some  of   the  zones---such   as  the   period-8  and   the  period-5
regions---are truncated at certain values of $\lambda$ and $\epsilon$.
Between the zones of  period-4 orbits and two-band  chaotic evolution,
higher-order periodic orbits as well as chaotic evolution in more than
two  bands  are  missing.   The  period-3  stability window is instead
clearly detected, immersed in the one-band chaotic regime.

Since  blurring  of  boundaries  between  zones  and  suppression   of
higer-order periodic and  chaotic orbits are  a direct consequence  of
the  internal  noise,  it  is  expected that---if noise decreases upon
incresing the number of  elements in the ensemble---the  phase diagram
becomes more sharply defined for higher $N$. It is however known that,
in  synchronous  globally  coupled  maps,  the internal noise does not
decrease in the usual self-averaging way as $N^{-1/2}$---as  predicted
by the so-called  law of large  numbers---but in a  much slower manner
\cite{kaneko90b,perez92a}. In  Section IV  we analyze  this aspect for
asynchronous  evolution,  finding  that  the  law  of large numbers is
recovered in these systems.

Note finally that if, in contrast with Fig. \ref{bif}, the bifurcation
diagram  would  be  plotted  for  fixed  $\epsilon$  as  a function of
$\lambda$, the bifurcation cascade  would proceed forwards. For  large
values of the  coupling intensity, however,  higher-order bifurcations
and chaos would not be reached even for $\lambda=4$.

\section{Mean field approach}

It  is  possible  to  describe  analytically  some  features  of   the
bifurcation cascade by performing  a kind of mean  field approximation
to  Eq.   (\ref{async}).  From  the  viewpoint  of  a  single element,
updating of its state occurs, in  average, once per time step. At  the
successive times where a given   element is updated, the sequence   of
values of the mean field  $F(t) = N^{-1} \sum_j f[x_j(t)]$  fluctuates
due to  the evolution  of the  individual states  of all the elements,
effectively resembling  a stochastic  process. In  such a  way,    the
single-element dynamics can be though  of as given by a  deterministic
map  subject  to  the  action  of an effective ``external'' stochastic
forcing.

In  order  to  characterize  this  effective  forcing,  we assume---as
suggested by  the numerical  simulations---that  the evolution  of the
system (\ref{async}) determines, at long times, a well-defined measure
$\mu(x)$ on the space $x$. At any time step the value of the  elements
will distribute according to  this measure and, for  $N\to\infty$, the
mean field $F(t)$ will approach a constant $F_0 = \int f(x)\mu(x) dx$.
For finite $N$, however, the field $F(t)$ will fluctuate around $F_0$,
so that we can write
\begin{equation}
\lim_{t\to\infty}F(t) \approx F_0 + \xi(t),
\label{xi}
\end{equation}
where $\xi(t)$ is a stochastic process of zero mean.

Within this approximation, the dynamics  of a single element is
\begin{equation} \label{noise}
x_i(t+1)= (1-\epsilon)f[x_i(t)]+\epsilon F_0 + \epsilon \xi(t),
\end{equation}
The effective individual  evolution is thus  given by a  deterministic
renormalized map $f^{\rm eff}(x)= (1-\epsilon)f(x)+\epsilon F_0$  upon
which an additive noisy force of zero mean acts.  It has to be noticed
that the constant $F_0$, being determined by the collective  dynamics,
prevents the population from complete decoupling.

It is well known that a  deterministic map subject to the action  of a
moderate additive noise of zero mean preserves its critical  behavior,
the  only  effect  of  noise  being  the  suppression  of   high-order
bifurcations. Within this picture,  we can then explicitely  calculate
the first bifurcation points as follows.

Suppose firstly that the map $f^{\rm eff}(x)$ has a stable fixed point
$x_0$ as its only attractor. Since---in the absence of noise---all the
elements  are  attracted  to  it,  we have $\mu(x)=\delta (x-x_0)$ and
$F_0=f(x_0)$.   The  effective  single-element  evolution becomes thus
$x_i(t+1) = (1-\epsilon) f[x_i(t)] + \epsilon f(x_0)$. The equation
\begin{equation}
x_0 = (1-\epsilon) f(x_0) +\epsilon f(x_0) = f(x_0),
\end{equation}
determines $x_0$, which  results to coincide  with the fixed  point of
the original  map $f(x)$.  The evolution  equation can  be now readily
linearized to  find that  $x_0$ is  stable if  the coupling  intensity
satisfies
\begin{equation}
\epsilon > \epsilon_c^1 = 1-|f'(x_0)|^{-1}.
\label{ecritic}
\end{equation}

For  the  logistic  map,  that  has  a  stable fixed point at $x_0=1 -
\lambda^{-1}$ for $1<\lambda<3$, Eq. (\ref{ecritic}) implies that  the
state $x_i=x_0$ is  stable if $\epsilon>1-|2-\lambda|^{-1}$.  That is,
for $2<\lambda<3$  the state  is stable  for all  $\epsilon$ and,  for
$\lambda>3$  it  is  stable  for  $\epsilon$  large  enough.  In  Fig.
\ref{phases}  we  have  drawn  the  curve $\epsilon_c^1(\lambda)$ that
defines the boundary between the fixed point and the period-2 regions.
This analytical result is in full agreement with simulations.

A  similar  argument  can  be  applied  to period-2 orbits. Due to the
randomness of the updating scheme,  at each time step practically  one
half of the elements will be at   one of the two states of the  orbit,
say $x_A$, and the other  half at the other, $x_B$.  The corresponding
invariant measure  is $\mu(x)  = [\delta(x-x_A)+\delta(x-x_B)]/2$  and
$F_0=[  f(x_A)+f(x_B) ]/2$. Therefore,  the  equations which determine
the values of $x_A$ and $x_B$ are
\begin{equation}
\begin{array}{rcl}
x_A &=& (1-\epsilon)f(x_B)+\epsilon[f(x_A)+f(x_B)]/2 \\
x_B &=& (1-\epsilon)f(x_A)+\epsilon[f(x_A)+f(x_B)]/2.
\end{array}
\label{xaxb}
\end{equation}
These equations  are equivalent  to $x_A=f(x_B)$  and $x_B=f(x_A)$, so
that  $F_0=(x_A+x_B)/2$  and  $x_{A,B}$   are  also  the  two   states
corresponding to the period-2 orbit  of the original map $f(x)$.   For
the logistic  map these  equations can  be solved,  and the  stability
condition implies
\begin{equation}
\epsilon>\epsilon_c^2=\frac{-1-2\lambda+\lambda^2-
\sqrt{1-6\lambda+3\lambda^2}}{\lambda(\lambda-2)}.
\label{ecritic2}
\end{equation}
The line $\epsilon_c^2(\lambda)$ is also shown in Fig.   \ref{phases},
separating the regions of period-2 and period-4 orbits. This result is
again in good agreement with simulations, though a slight deviation is
observed for large values  of $\epsilon$, i.e. near  $\lambda=4$. This
deviation  can  be  ascribed  to  the  increasing  effect of noise for
growing $\epsilon$ (cf. Eq. (\ref{noise})), that blurs the  boundaries
where bifurcations occur.

Analogous reasoning may  be used to  determine the boundaries  between
other zones. For higher-order periodic orbits, however, the  equations
cannot  be  explicitly  solved  in  the  case of the logistic map. For
chaotic evolution, moreover, the  measure $\mu(x)$ should be  obtained
numerically.

\section{Fluctuations of the mean field}

A key ingredient in the analysis presented in the previous section  is
the assumption  that, for  large populations, the system  (\ref{async})
defines a measure  in the one-element  state space $x$,  such that for
sufficiently  long   times  the   elements  exhibit   a   well-defined
distribution  in  $x$.   It  is  implicit  in that assumption that the
amplitude  of  internal  noise  decreases  as  $N$ grows. As mentioned
above,  however,  previous  work   has  shown  that  fluctuations   in
deterministic   synchronous   globally   coupled   systems   does  not
self-average  in  the  usual  way  \cite{kaneko90b,perez92a}.   It  is
therefore worthwhile  to analyze  this point  in some  detail for  the
present case of asynchronous update.

The fluctuating part of the  mean field, $\xi(t)$ in Eq.   (\ref{xi}),
can also be analytically  studied within some further  approximations.
In order to illustrate our arguments, let us restrict ourselves to the
period-2 orbit, where $\mu(x) = [\delta(x-x_A)+\delta(x-x_B)]/2$.

Due to the fluctuations arising at finite values of $N$, the number of
elements in  each state  will not  be $N/2$,  but rather a fluctuating
number $n_A(t)$  at $x_A$  and $n_B(t)=N-n_A(t)$  at $x_B$.  Since the
mean  field  $F(t)$  will  consequently  differ  from  $F_0= [f(x_A) +
f(x_B)]/2 = (x_A+x_B)/2$, this implies that the individual states will
spread around the values $x_A$ and $x_B$. At a given time,  therefore,
the elements near $x_{A,B}$ will have states $x_{i_{A,B}} = x_{A,B}  +
\delta  x_{i_{A,B}}  (t)$.  The  fluctuating  mean  field will thus be
\begin{equation}
F(t) = {1\over N}\left[ \sum_{i_A} f(x_A+\delta x_{i_A})+
\sum_{i_B} f(x_B+\delta x_{i_B}) \right]
\end{equation}
where the sums $\sum_{i_A}$ and $\sum_{i_B}$ run over the elements near
$x_A$ and $x_B$, respectively.

In the  lowest-order approximation,  we have  $F(t) =  [n_A f(x_A)+n_B
f(x_B)] /N=(n_A x_B+n_B x_A)/N$ that can be rewritten as
\begin{equation}
F(t)= \frac{x_A+x_B}{2}+{1\over N} \left[
n_A(t)-\frac{N}{2}\right](x_B-x_A).
\label{fluct}
\end{equation}
In this expression we identify  the fluctuations of $F(t)$ as  $\xi(t)
\equiv [n_A(t)-N/2](x_B-x_A)/N$. Since  updating is applied  at random
in the ensemble, $n_A(t)$ is expected to have a binomial  distribution
around  $N/2$.   This  readily  shows  that  $\xi (t)$ is a stochastic
process  with  zero  mean,  whose  mean square dispersion $\Delta \xi$
depends on $N$ as $\Delta \xi  \sim N^{-1/2}$.  We also note  that the
amplitude  of   fluctuations  depends   linearly  on   the  difference
$|x_A-x_B|$  between  the  two  states  of  the  period-2  orbit. As a
byproduct,  this  indicates  that  the  fixed-point  state exhibits no
fluctuations. In Fig. \ref{fluct2} we  have plotted $\Delta \xi$ as  a
function of $|x_A-x_B|$ for a fixed value of $\lambda$ in a population
of $10^4$ logistic maps.  The coupling intensity $\epsilon$ runs  over
the  region  of  period-2  orbits,  making $|x_A-x_B|$ vary from small
values  near  $\epsilon=\epsilon_c^1$  to  larger values as $\epsilon$
decreases.  A large region in which $\Delta \xi$ varies linearly  with
$|x_A-x_B|$ is clearly observed.

\begin{figure}[t]
\centerline{
\resizebox{\columnwidth}{!}{\rotatebox{-90}{\includegraphics{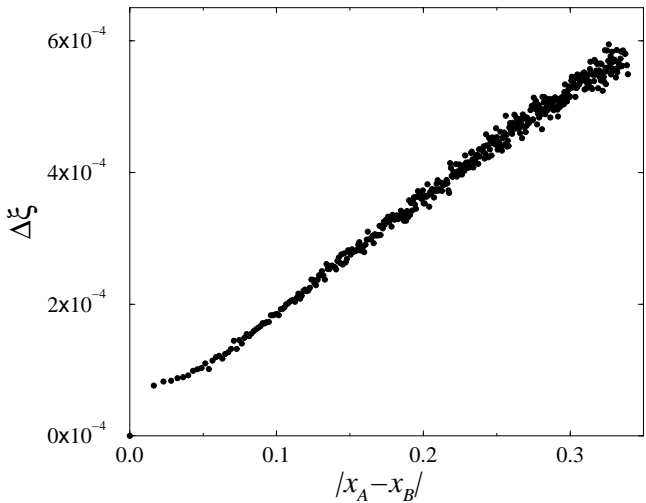}}}}
\caption{The standard deviation of the fluctuations of the mean field,
$\Delta \xi$,  as a  function of  the the  separation between  the two
states  of  the  period-2  orbit.  The  system  size  is  $N=10^4$ and
$\lambda=3.4$; $\epsilon$ approaches  $\epsilon_c^1$ from below.   The
abcisas used to plot  the data correspond to  the values of $x_A$  and
$x_B$  provided  by  Eq.    (\protect\ref{xaxb})  as  a  function   of
$\epsilon$.}
\label{fluct2}
\end{figure}

Remarkably, the  proportionality of  the mean  field fluctuations with
$N^{-1/2}$  is  also  numerically  verified  in  the regions with more
complex dynamics. The validity of the law of large numbers proves thus
to  be  a  generic  feature  in  the  fluctuations of the asynchronous
ensemble of  globally coupled  maps. In  Fig. \ref{fluct1}  we show  a
double logarithmic plot of the mean square deviation $\Delta F$ of the
mean field  as a  function of  the system  size for $\epsilon=0.1$ and
$\lambda=4$, i.e. in the chaotic regime. The $N^{-1/2}$-dependence  is
apparent.

\begin{figure}[h]
\centerline{
\resizebox{\columnwidth}{!}{\rotatebox{-90}{\includegraphics{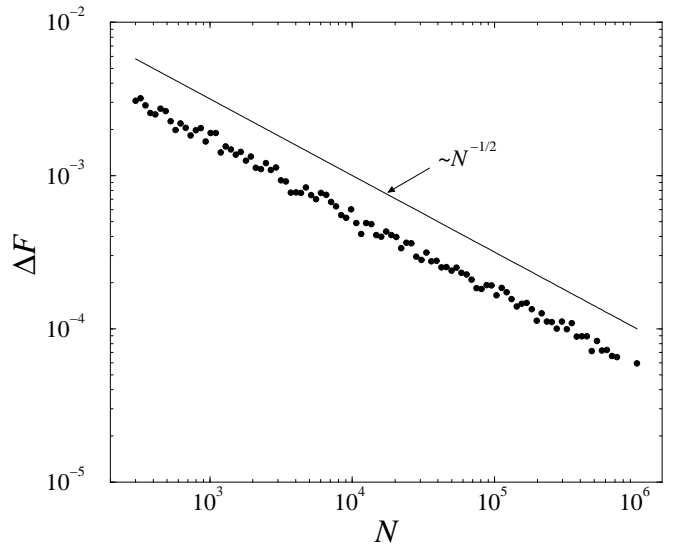}}}}
\caption{The standard deviation of the mean field, $\Delta F$, as  a
function of  the system  size. The  double logarithmic  plot shows the
decaying behavior $N^{-1/2}$ for system sizes ranging from $N=300$  to
$10^6$ for $\lambda =4$ and $\epsilon  = 0.1$. A line of slope  $-1/2$
is also shown for comparison.}
\label{fluct1}
\end{figure}

\section{Mean field dependence on the coupling intensity}

In order to  complete our discussion,  it is worthwhile to investigate
how  the  collective  behavior  varies  as  the  coupling intensity is
modified.   In  this  section  we  present  numerical  results  on the
dependence of $F(t)$ on $\epsilon$  for a given value of  $\lambda$ in
an ensemble of logistic maps.

In  Figure  \ref{meanfield}  we  show  the  time-averaged  mean  field
$\langle  F\rangle$  as  a  function  of  the  coupling intensity, for
$\lambda=4$ and $N=10^3$. At $\epsilon=0$ the elements are  completely
uncoupled. In the fully  chaotic regime ($\lambda=4$), the  individual
states are symmetrically distributed in the interval $(0,1)$ according
to the invariant measure of the logistic map. This uncorrelated  state
is then characterized by an average field $ \langle F \rangle =  0.5$.
For nonzero values  of the coupling  intensity $\epsilon$, we  observe
that a correlated state develops, characterized by values of  $\langle
F\rangle$ larger than $0.5$. The average mean field grows however in a
highly nonmonotonic way. Large  downward peaks can be  observed, where
the collective  state again  approaches the  value $\langle  F \rangle
=0.5$. The  broadest of  these peaks,  near $\epsilon=0.07$, coincides
with the period-3 stability window.  The inset of Fig. \ref{meanfield}
shows a detail of  the same function in  the region of small  coupling
intensity (for an ensemble of  $5\times 10^4$ elements).  The  average
mean field  exhibits a  complex dependence  on $\epsilon$,  seemingly,
with self-similar features. These features are probably reflecting the
presence of higher-order stability windows in the chaotic regime.  For
large values of $\epsilon$ the system abandons the chaotic state  and,
for $\epsilon\approx 0.2$,  it enters the  regime of periodic  orbits.
Correspondingly, $\langle F\rangle$ behaves less erratically.

\begin{figure}[t]
\centerline{
\resizebox{\columnwidth}{!}{\rotatebox{-90}{\includegraphics{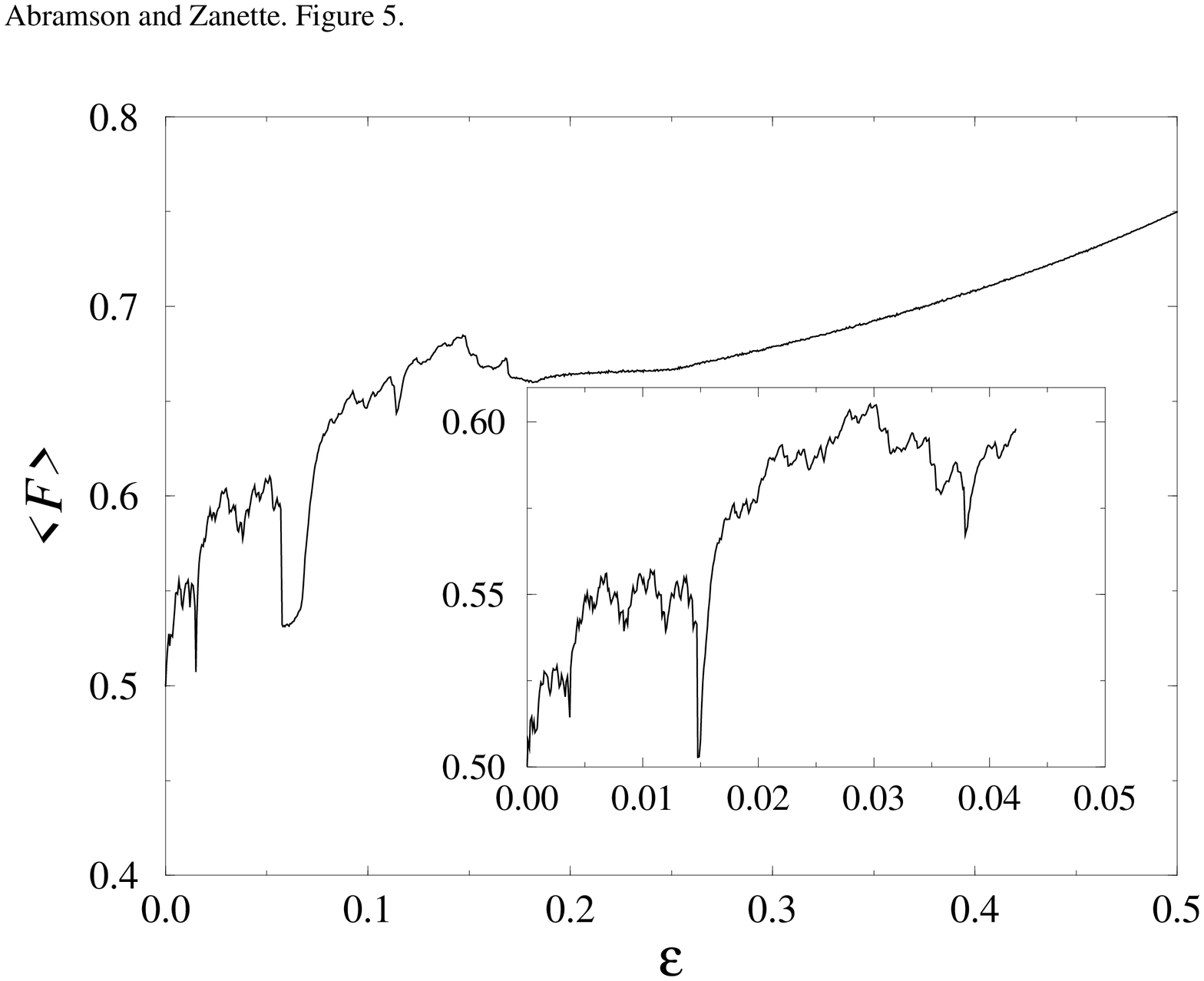}}}}
\caption{Temporal  average  of  $F$  as  a function of $\epsilon$. The
system has $N=1000$ elements  and $\lambda=4$. Inset: a  detailed view
of the region of small $\epsilon$, with $N=50000$.}
\label{meanfield}
\end{figure}

For lower values  of $\lambda$ a  similar picture can  be observed. In
such cases,  of course,  the average  mean field does not  start at $F
=0.5$ for  $\epsilon=0$, since  the values  of $x_i$  do not cover the
interval  $(0,1)$  with  a  symmetric  distribution.   As the coupling
intensity  grows,  at  the  critical  value $\epsilon_c^1$, the entire
population  collapses  into  the  fixed  point  and $\langle F\rangle$
reaches a fixed maximal value.

Finally, we  have studied  how the  fluctuations of  $F(t)$ around its
average value $\langle F \rangle$ depend on the coupling intensity. In
Figure \ref{peaks} we have plotted the mean square dispersion of  $F$,
$\Delta  F  =  \sqrt{\langle  (F-\langle  F\rangle  )^2\rangle}$, as a
function of $\epsilon$, for several system sizes.  These plots show  a
rather uniform background, that disappears at $\epsilon=0.5$ when  the
whole system is attracted to the fixed point (the non-linear parameter
is $\lambda=4$). This  fluctuation level is  reduced by enlarging  the
system. Superimposed to this background, some sharp spikes of enhanced
fluctuations are  seen for  low values  of $\epsilon$.   They coincide
with   the   downward   peaks   of   $\langle   F\rangle$   in  Figure
\ref{meanfield},  and  correspond  to  the  stability  windows  in the
chaotic regime.  As illustrated  in the inset of Fig.  \ref{peaks} for
the  widest  period-3  window,  the  behavior  of  any  element of the
ensemble in these regions is highly intermittent. During certain  time
intervals,  the  elements   are  engaged  in   periodic  orbits   but,
occasionally,  they  exhibit  a  regime  of  chaotic  evolution.  This
intermittency  between  two  qualitatively  different forms of motion,
each of them having specific values of $\langle F\rangle$ and  $\Delta
F$, causes the overall mean-field dispersion to attain unusually large
levels.  These  anomalous fluctuations can  even grow upon  increasing
the system size.

\begin{figure}[b]
\centerline{
\resizebox{\columnwidth}{!}{\rotatebox{-90}{\includegraphics{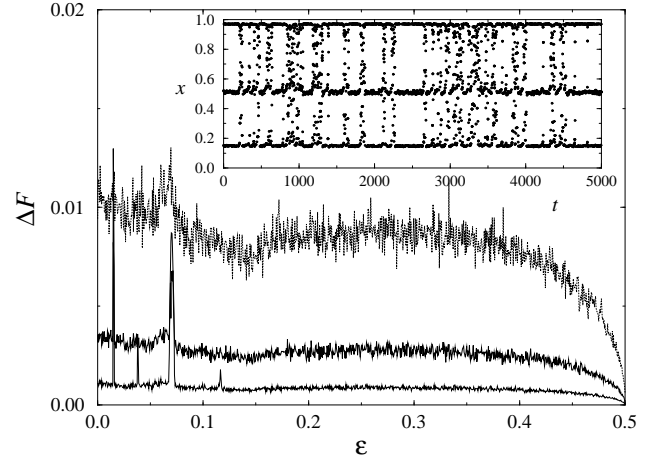}}}}
\caption{The  mean  square  dispersion  $\Delta  F$  as  a function of
$\epsilon$,  for  several  system  sizes.  Dotted line: $N=10^3$; full
thick line: $N=10^4$;  full thin line:  $N=10^5$. Increasing $N$,  the
background is seen to diminish, while the peaks are seen to grow.  For
$N=1000$, the peaks cannot be resolved from the background. Inset: The
orbit of  a single  element at  $\lambda=4$ and $\epsilon=0.069$, from 
a system with 1000 elements, displaying intermittent  behavior between
a period-3 orbit and chaotic motion.}
\label{peaks}
\end{figure}

\section{Summary and Conclusions}

In this paper we have analyzed the collective behavior of an  ensemble
of globally  coupled maps  whose dynamics  is updated  asynchronously.
Asynchronous  updating,  whose  study  is  motivated  by  the  aim  at
realistically modelling  the evolution  of real  systems, introduces a
stochastic ingredient in the dynamics, with nontrivial consequences in
the  behavior  of  the  population.  Although  some  of our analytical
results  hold  for  any  kind  of  coupled  maps,  we have focused our
attention---in particular, in the numerical simulations---in the  case
of logistic maps, $f(x)=\lambda x(1-x)$.

We have  numerically found  that, for  a fixed  value of the parameter
$\lambda$,  increasing  the  coupling  intensity  leads  the system to
simpler and simpler evolution, running backwards over the  bifurcation
diagram of the logistic map. For sufficiently large coupling, in fact,
a  fixed-point  state  is  reached  for  the whole ensemble, even when
$\lambda$ corresponds to chaotic individual motion. This is  in strong
contrast with the behavior  of globally coupled maps  with synchronous
updating. Indeed, large  coupling intensities lead  such systems to  a
synchronized state where all the elements reproduce the evolution of a
single,  uncoupled  map.  Below  the synchronization threshold, those
systems exhibit  a regime  of clustering  not observed  in the present
case of asynchronous updating.

An approximate  analytical description  for the  asynchronous ensemble
can be achieved by constructing an effective single-element  dynamics,
which  result  to  be  driven  by the map $f^{\rm eff}(x)=(1-\epsilon)
f(x)+\epsilon F_0$,  where $\epsilon$  is the  coupling intensity  and
$F_0$  is  a  constant  to  be  determined  selfconsistently  from the
collective evolution. For the case of logistic maps, this  approximate
picture makes it possible  to analytically calculate the  threshold of
the lower-order bifurcations.  These results compare successfully with
numerical simulations. Note that  as $\epsilon$ increases, the  weight
of the nonlinear term in the effective dynamics decreases,  explaining
why  the  bifurcation  diagram  develops  backwards  when the coupling
intensity grows.

The   effective   individual   dynamics   is   affected   by  internal
fluctuations, which enter the single-element evolution as an  additive
noise term. As is well known for noisy maps, the main effect of  these
fluctuations consists of  suppression of higher-order  bifurcation and
blurring of both regular and  chaotic motion. We have studied  how the
amplitude of the internal noise depends on the ensemble size, i.e.  on
the  number  $N$  of  elements  in  the  population,  by analyzing the
temporal behavior of the mean field $F(t) = N^{-1} \sum_i  f[x_i(t)]$.
Numerical simulations show that  the fluctuations of $F(t)$  decreases
with the system size as  $N^{-1/2}$. This is again in  strong contrast
with synchronous  globally coupled  maps, for  which it  is known that
fluctuations  decrease  in  a  much  slower  manner.  Here,   instead,
the   intrinsic   stochastic   character   of   the   evolution  makes
fluctuations to  obey the  usual selfaverage  statistics, and  the law
of  large  numbers  holds.  This  result  suggest  that  the violation
of  the  law  of   large  numbers  is  not   a  robust  feature   upon
introduction of random  elements in the  dynamics of globally  coupled
ensembles.

In summary, we have shown that   the  collective behavior of  globally
coupled maps with asynchronous updating exhibits important differences
when compared with that of  synchronous dynamics. Coupling is able  to
suppress  the   complexity  of   individual  evolution   and  internal
fluctuations  selfaverage  in  the  usual  way  as  the  system   size
increases. The extension of these results to ensembles formed by  more
complex maps  and by  continuous-time dynamical  systems should be the
subject of further analysis.


\end{document}